\documentclass[12pt,preprint]{aastex631}

\newcommand{\kms}{{km s$^{-1}$}}

\shorttitle{High-Velocity Scatterer}
\shortauthors{Leonard et al.}

\begin{document}

\title{A High-Velocity Scatterer Revealed in the Thinning Ejecta of
  a Type II Supernova}

\correspondingauthor{Douglas C. Leonard}
\email{dleonard@sdsu.edu}

\author[0000-0001-7839-1986]{Douglas C. Leonard}
\affiliation{Department of Astronomy, San Diego State University, San
  Diego, CA 92182-1221, USA}

\author[0000-0003-0599-8407]{Luc Dessart}
\affiliation{Institut d'Astrophysique de Paris, CNRS-Sorbonne Universit\'e,
  98 bis boulevard Arago, F-75014 Paris, France}

\author[0000-0001-5094-8017]{D. John Hillier}
\affiliation{Department of Physics and Astronomy \& Pittsburgh Particle
  Physics, Astrophysics, and Cosmology Center (PITT PACC), University of
  Pittsburgh, 3941 O’Hara Street, Pittsburgh, PA 15260, USA}

\author[0000-0003-0006-0188]{Giuliano Pignata}
\affiliation{Departamento de Ciencias Fisicas, Universidad Andres Bello,
Avda. Republica 252, Santiago, Chile}
\affiliation{Millennium Institute of Astrophysics (MAS), Nuncio Monsenor
  Sotero Sanz 100, Providencia, Santiago, Chile}

\author[0000-0002-3452-0560]{G. Grant Williams}
\affiliation{Steward Observatory, University of Arizona, 933 North Cherry
  Avenue, Tucson, AZ 85721-0065, USA}
\affiliation{MMT Observatory, P.O. Box 210065, University of Arizona,
  Tucson, AZ 85721, USA}

\author[0000-0003-1495-2275]{Jennifer L. Hoffman}
\affiliation{Department of Physics \& Astronomy, University of Denver, 2112
East Wesley Avenue, Denver, CO 80208, USA}

\author[0000-0002-0370-157X]{Peter Milne}
\affiliation{Steward Observatory, University of Arizona, 933 North Cherry
  Avenue, Tucson, AZ 85721-0065, USA}

\author[0000-0001-5510-2424]{Nathan Smith}
\affiliation{Steward Observatory, University of Arizona, 933 North Cherry
  Avenue, Tucson, AZ 85721-0065, USA}

\author[0000-0002-5083-3663]{Paul S. Smith}
\affiliation{Steward Observatory, University of Arizona, 933 North Cherry
  Avenue, Tucson, AZ 85721-0065, USA}

\author[0000-0002-9011-0216]{Harish G. Khandrika}
\affiliation{Department of Astronomy, San Diego State University, San
  Diego, CA 92182-1221, USA}
\affiliation{Space Telescope Science Institute, 3700 San Martin Drive,
  Baltimore, MD 21218, USA}

\begin{abstract}

We present deep, nebular-phase spectropolarimetry of the Type II-P/L SN~2013ej,
obtained 167 days after explosion with the European Southern Observatory's Very
Large Telescope.  The polarized flux spectrum appears as a nearly perfect
($92\% {\rm\ correlation}$), redshifted (by $\sim 4,000$ \kms) replica of the
total flux spectrum.  Such a striking correspondence has never been observed
before in nebular-phase supernova spectropolarimetry, although data capable of
revealing it have heretofore been only rarely obtained.  Through comparison
with 2D polarized radiative transfer simulations of stellar explosions, we
demonstrate that localized ionization produced by the decay of a high-velocity,
spatially confined clump of radioactive $^{56}$Ni --- synthesized by and
launched as part of the explosion with final radial velocity exceeding $4,500$
\kms\ --- can reproduce the observations through enhanced electron scattering.
Additional data taken earlier in the nebular phase (day 134) yield a similarly
strong correlation ($84\%$) and redshift, whereas photospheric-phase epochs
that sample days 8 through 97, do not.  This suggests that the primary
polarization signatures of the high-velocity scattering source only come to
dominate once the thick, initially opaque hydrogen envelope has turned
sufficiently transparent.  This detection in an otherwise fairly typical
core-collapse supernova adds to the growing body of evidence supporting strong
asymmetries across Nature's most common types of stellar explosions, and
establishes the power of polarized flux --- and the specific information
encoded by it in line photons at nebular epochs --- as a vital tool in such
investigations going forward.

\end{abstract}


\section{Introduction}
\label{sec1}

The mechanism that successfully ejects a star's outer layers in a core-collapse
supernova (CC SN) explosion remains uncertain \citep[for a recent overview,
  see][]{Ono20}. A powerful clue to its nature is the geometry of the ejected
material, and the observational technique most capable of directly revealing
this geometry for an unresolved source is polarimetry \citep{Shapiro82}. Under
the {\it ansatz} of simple reflection by an external scatterer that does not
present circular symmetry in the plane of the sky, the polarized flux --- an
object's observed polarization percentage multiplied by its total flux spectrum
--- reveals the spectrum of any light that has ``taken a bounce'' prior to
reaching the observer \citep{Hough06}.  This basic framework has been most
famously employed to establish the presence of hidden broad-line regions in
Type II Seyfert galaxies \citep{Antonucci85}.  Although not generally observed
in Seyfert studies, a pronounced redshift of the polarized flux relative to the
total flux demands one additional characteristic of the scattering source:
Radial recession from the region responsible for the total flux spectrum
\citep{Antonucci85,Kawabata02,Jeffery04,Dessart21b}.  In this study we present
and examine the first nebular-phase spectropolarimetry data of a CC SN,
SN~2013ej, to bear the hallmarks of such a high-velocity
scatterer.\footnote{Due to the faintness of the targets, nebular-phase CC SN
  polarization measurements have typically been limited to calculations of
  broadband averages rather than consideration of the full,
  wavelength-dependent spectropolarimetry being discussed here (see
  \citealt{Nagao19} for a recent compilation of existing nebular-phase SN
  data).}

SN~2013ej occurred in the nearby galaxy M74 ($D = 9.6\pm 0.7 {\rm\ Mpc}$;
\citealt{Huang15}), and has been extensively studied at X-ray, UV, optical, and
infrared wavelengths (see \citealt{Morozova18}, and references therein).
Photometrically, it straddles the boundary between the Type II-Plateau (II-P)
and Type II-Linear (II-L) sub-classifications (\citealt{Dhungana16}; see also
Figure~\ref{plot103b} in the Appendix), and exhibits characteristics that
suggest some early interaction with circumstellar material (CSM;
\citealt{Morozova18,Hillier19}).  From tight constraints placed by
pre-explosion imaging, \citet{Valenti14} estimate the date of explosion to have
been $23.95^{+4{\rm\ h}}_{-12{\rm\ h}}$ July, 2013 UT.  \citet{Yuan16} identify
the point of transition between the ``photospheric'' (opaque) and ``nebular''
(increasingly transparent) phases of its development to have occurred around
day 106 after explosion.

SN~2013ej's spectroscopic evolution was fairly typical for an SN~II-P/L through
the end of the photospheric phase \citep{Yuan16}.  As is commonly observed
\citep{Silverman17}, its nebular spectra revealed marked asymmetries in
prominent line profiles (e.g., H$\alpha$; \citealt{Yuan16}), a characteristic
often ascribed to the ionizing effects of $^{56}$Ni decay occurring farther out
into the ejecta than conventional explosion mechanisms generally predict
\citep[see][and references therein]{Ono20}.  In the case of SN~2013ej,
\citet{Utrobin17b} propose the ejection of $^{56}$Ni at velocities exceeding
$4,000$ \kms\ in an effort to explain both the nebular line-profile morphology
as well as its bolometric luminosity evolution.  Such high inferred velocities
strain the maximum values obtainable from standard neutrino-driven CC SN
simulations \citep{Wongwathanarat15,Orlando21}.

SN~2013ej exhibited strong polarization during the photospheric phase, and
there have now been two thorough spectropolarimetric studies carried out on its
early evolution.  In the first, \citet{Mauerhan17} find a non-spherical
photosphere, perhaps induced by weak interaction with an asymmetric CSM to be
the probable causes of the photospheric-phase polarization; reflection off of
pre-existing, distant dust is also considered as a possible contributor to the
signal.  In the second study, \citet{Nagao21} invoke a mix of explosion
asymmetry and CSM interaction as the causes of the continuum polarization seen
during the photospheric phase.

In this {\it Letter}, we focus exclusively on high signal-to-noise ratio (SNR)
spectropolarimetry of SN~2013ej obtained during the nebular phase, in an effort
to uniquely constrain the scattering environment at late times. We describe the
observations in Section~\ref{sec2}, present the results and analysis in
Section~\ref{sec3}, and conclude in Section~\ref{sec4}.  Appendices provide
additional detail of the data acquisition and reductions
(Appendix~\ref{appendixa}), empirical analysis techniques
(Appendix~\ref{appendixb}), and our modeling approach and the inferences drawn
from it (Appendix~\ref{appendixc}).

\section{Observations}
\label{sec2}

We observed SN~2013ej on the nights of January 3, 4, and 12, 2014 for a total
of three hours using the European Southern Observatory's 8-meter Very Large
Telescope (VLT) with the FOcal Reducer and low dispersion Spectrograph 2
(FORS2; \citealt{Appenzeller98}) + polarimetry module.\footnote{VLT observing
  program 091.D-0401(A): PI Pignata.}  These were the final observations of a
multi-epoch campaign (see Appendix~\ref{appendixa1}) to gather
spectropolarimetry of this nearby SN.  They took place more than 160 days after
the estimated date of explosion and about 60 days after the photospheric phase
is estimated to have ended.  The total flux spectra exhibit all of the
characteristics of a nebular spectrum according to the criteria of
\citet{Silverman17}, indicating an optically thin, largely transparent nebula.

The observed polarization did not measurably change over the course of the nine
days of our observations, which allowed us to combine the datasets to improve
the SNR.  We shall refer to these combined observations as having occurred on
``Day 167'' after explosion, which represents the average of the epochs
corresponding to the three individual nights.  We corrected these data for an
interstellar polarization (ISP) component ($p_{\rm ISP} \approx 0.8\%$; see
Appendix~\ref{appendixa2}) as well as a small degree of instrumental
polarization ($p_{\rm inst} < 0.1\%$; \citealt{Cikota17}).

\section{Results and Analysis}
\label{sec3}

We compare the observed polarized and total flux spectra of SN~2013ej in
Figure~\ref{plot307f7a}a.  The two spectra show a remarkable resemblance, but
with an evident offset between them, in the sense that the polarized flux
appears shifted to longer wavelengths relative to the total flux.  To quantify
this, we cross-correlated the two spectra over the entire spectral range ($4300
- 9100$ \AA) using the FXCOR package within IRAF, which implements the
cross-correlation technique developed by \citet{Tonry79}. It yielded a single,
clear peak at $3,946 \pm 132 {\rm\ km\ s}^{-1}$, and reported a 92\%
correlation and \citeauthor{Tonry79} $R$ value of 14.6, both of which indicate
an ``excellent'' fit (Appendix~\ref{appendixb1}).  We compare the polarized and
total flux spectra --- now with the total flux spectrum artificially redshifted
by $3,946 {\rm\ km\ s}^{-1}$ --- in Figure~\ref{plot307f7a}b.  Such a
redshifted, polarized flux is the generic signature of scattering by a rapidly
receding, external source (see Section~\ref{sec1}); to help fix ideas, a
simplified model of the situation is presented in Figure~\ref{plot309f2}.  We
pursue this basic model and its implications now by confronting it with the
following three observational facts:

\begin{figure}[!htbp]
\centering
\rotatebox{90}{
 \scalebox{1.6}{
\includegraphics[width=3.4in]{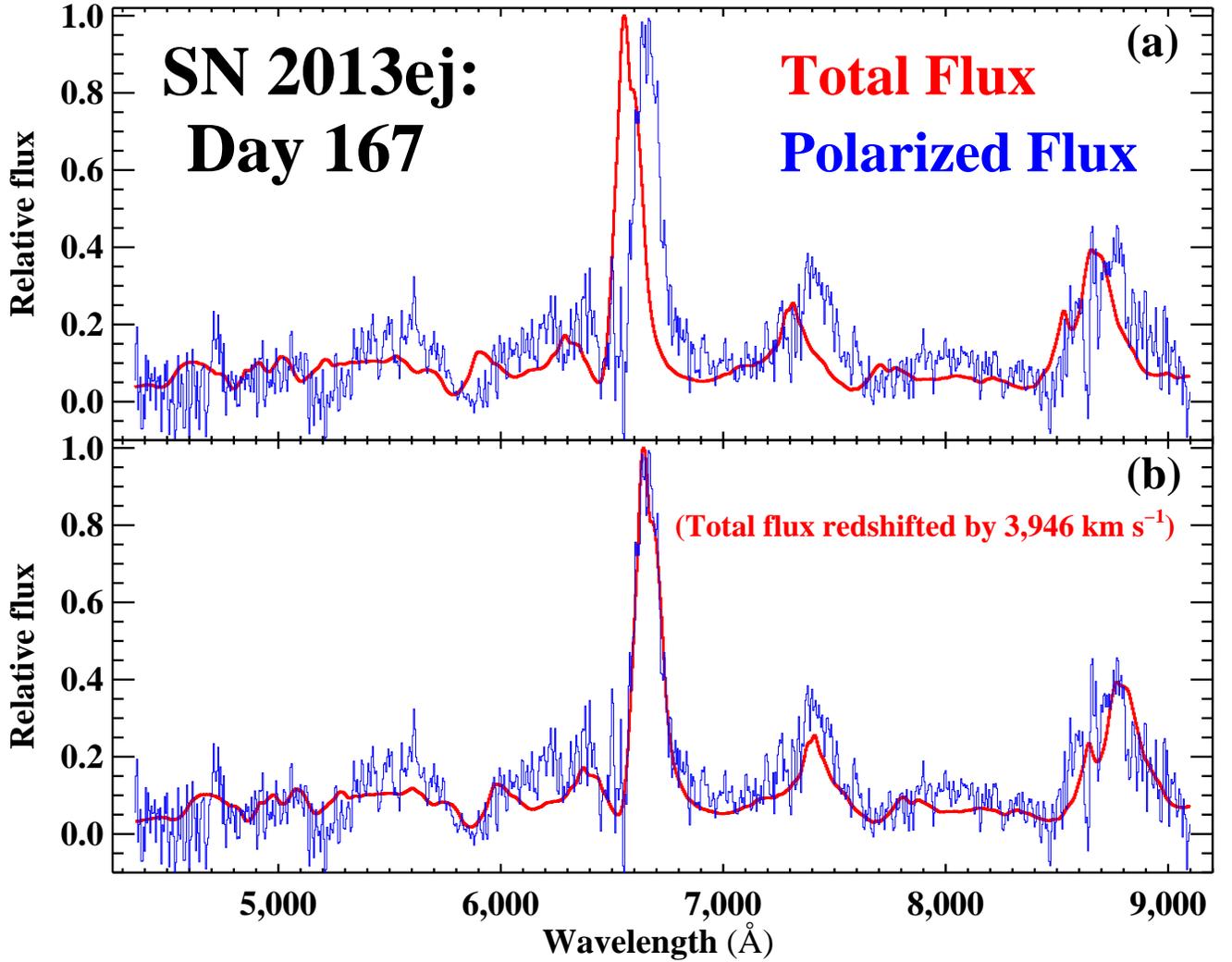} }}
 \caption{Polarized flux (blue, noisier data) and total flux (red, thick lines)
   spectra of SN~2013ej, taken 167 days after explosion. The recession velocity
   of $657$\ \kms\ \citep{Lu93} has been removed from these and all data unless
   otherwise noted.  The polarized flux was derived by multiplying the total
   flux by the observed polarization percentage at each wavelength, where the
   polarization percentage was estimated through calculation of the rotated
   Stokes parameter (RSP; see Appendix~\ref{appendixa1}).  (a) Observed data.
   (b) Same as in (a) but with the total flux spectrum now artificially
   redshifted by $3,946$ \kms.  To aid comparison of features, the total flux
   spectrum was first scaled by a factor of 0.0041 (chosen to produce the best
   fit to the H$\alpha$ spectral region), and then normalized to unity with the
   same normalization factor also applied to the polarized
   flux. \label{plot307f7a}}
\end{figure}

\begin{figure}[!htbp]
\begin{center}
 \hspace*{-0.3 cm}
\rotatebox{270}{
 \scalebox{1.5}{
 \includegraphics[width=3.4in]{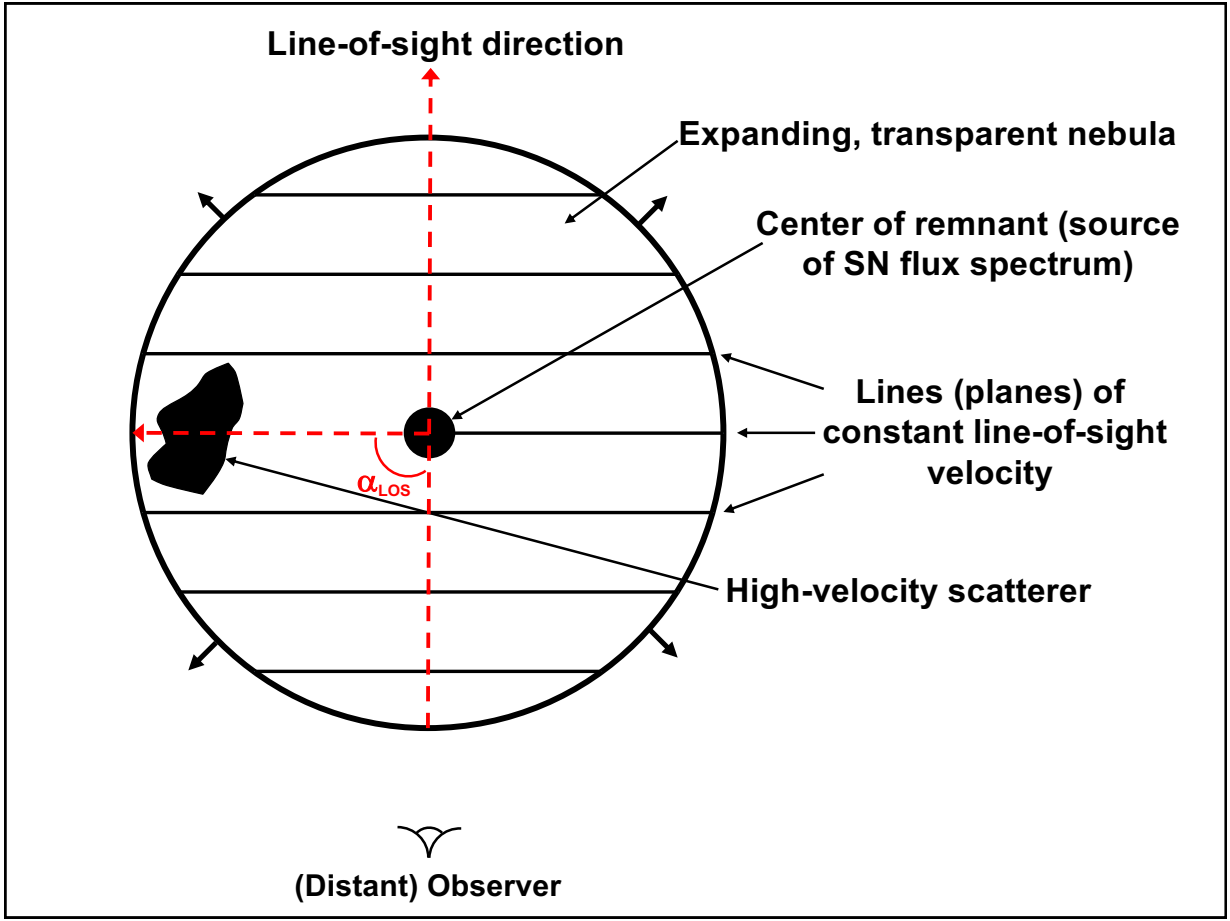}
}} 
 \caption{ Simplified schematic of physical situation proposed to generate the
   redshifted polarized flux spectrum seen in SN~2013ej at nebular epochs.  The
   SN flux spectrum is produced interior to an external scatterer that is
   located within the (homologously) expanding and otherwise transparent nebula
   that makes up the SN remnant.  The scattering source recedes from the center
   of the nebula with velocity $v_{\rm rec}$ at an angle to the line-of-sight
   (LOS), chosen here to be $\alpha_{\rm LOS} = 90^\circ$. This yields a LOS
   velocity relative to the distant observer of $v_{\rm LOS} = v_{\rm
     rec}\cos{\alpha_{\rm LOS}}$.  Under this arrangement, the portion of the
   total flux spectrum that is scattered by the source and redirected towards
   the observer receives both a net polarization and a redshift with respect to
   the total flux spectrum received directly.  The polarization arises from the
   scattering process (i.e., $p \propto \sin^2[\alpha_{\rm LOS}]$, achieving a
   maximum for $\alpha_{\rm LOS} = 90^\circ$), while the redshift is imparted
   according to $v_{\rm shift} = v_{\rm rec} - v_{\rm LOS} = v_{\rm rec}(1 -
   \cos{\alpha_{\rm LOS}})$.  Note that for simplicity the SN flux spectrum is
   depicted here as being generated solely in the innermost part of the ejecta;
   in reality (and, in our models), line emission occurs through an extended
   ejecta volume, resulting in some departures from a perfect replication of
   the total flux in the polarized flux spectrum (see
   Appendix~\ref{appendixc1}).}
   \label{plot309f2}
\end{center}
\end{figure}

\begin{figure}[!htbp]
\centering
\rotatebox{90}{
 \scalebox{1.6}{
 \includegraphics[width=3.4in]{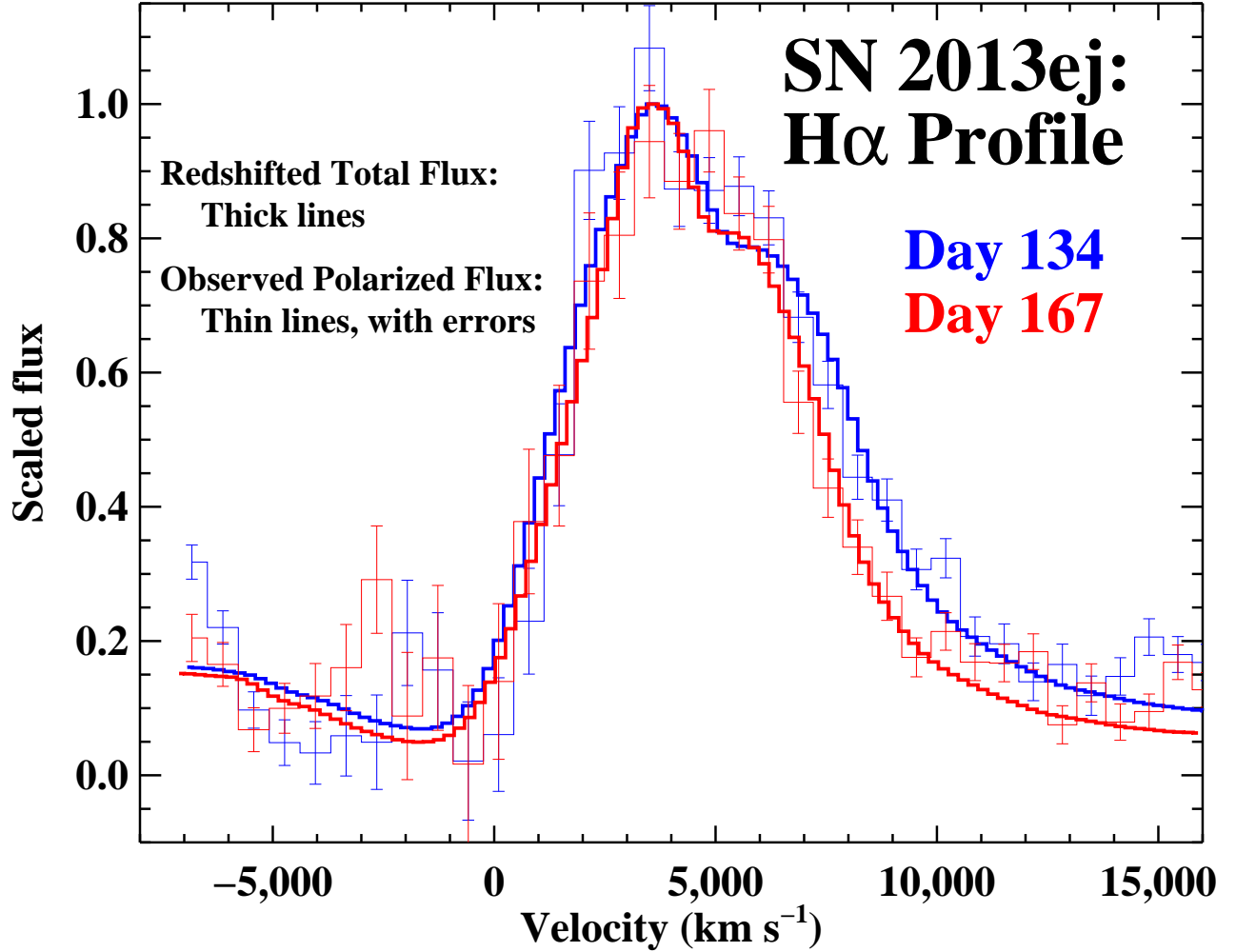} }}
 \caption{Polarized flux (thin lines, with 1$\sigma$ statistical uncertainties)
   and total flux (thick lines) spectra across the H$\alpha$ profile of
   SN~2013ej during the two nebular epochs sampled by our observations.  The
   total flux spectra have been artificially redshifted by $4,368$ \kms\ (Day
   134) and $3,946$ \kms\ (Day 167), and scaled by factors of 0.0035 (Day 134)
   and 0.0041 (Day 167), to assist with the direct comparison of features.  The
   peaks of the total flux at both epochs have been normalized to unity, with
   the same scaling then also applied to the polarized flux.  Note that the
   polarized flux data have a wider binning than the total flux data to improve
   the SNR. \label{plot304f3}}
\end{figure}

\begin{enumerate}

\item The first clear signature of the high-velocity scatterer emerged sometime
  between days 97 and 134 after explosion.  We searched all other epochs of our
  data (i.e., those from days 8, 34, 55, 67, 97, and 134; see Table~\ref{tab:1}
  in the Appendix) for significant correlation between the total and polarized
  fluxes by running FXCOR at each epoch in a manner identical to our treatment
  of the day 167 data.  The other nebular-phase epoch, taken on day 134,
  yielded results very similar to, albeit with somewhat less significance than,
  those obtained on day 167: A single, strongly peaked correlation at $v_{\rm
    shift} = 4,368 \pm 232$ \kms, Tonry \& Davis $R$ value of 7.6, and a
  correlation percentage of 84\%.  None of the earlier, photospheric epochs
  yielded a Tonry \& Davis $R$ value greater than 1.7 or a correlation greater
  than 34\% for any velocity shift (see Table~\ref{tab:2} in the Appendix).

\item The total and polarized flux profiles of strong emission lines (i.e.,
  H$\alpha$ $\lambda$6563 \AA; [\ion{Ca}{2}] $\lambda\lambda$7291, 7324 \AA;
  \ion{Ca}{2} $\lambda\lambda$8498, 8542, 8662 \AA) are closely matched in
  terms of breadth and specific spectral characteristics including,
    notably, asymmetries in the line profiles.  We show the data for the
  H$\alpha$ region in Figure~\ref{plot304f3} for both nebular epochs.

\item The overall spectral energy distributions (that is, the spectral shapes
  across the observed range from $4300$ \AA\ --- $9100$ \AA) of the total and
  polarized fluxes are the same (Figure~\ref{plot307f7a}, and
  Figure~\ref{plot307f6a} in the Appendix).

\end{enumerate}

\noindent Taken in turn, these facts generically imply that the scattering
source must (1) have first come to dominate the polarization signal during the
early nebular phase of the SN's development; (2) possess some degree of
geometrical confinement and be situated largely external to the region
responsible for the bulk of the line emission; and (3) be effectively ``gray''
(that is, independent of wavelength) in its scattering opacity.  This final
characteristic argues naturally for free electrons as the scattering
agent.\footnote{See Appendix~\ref{appendixc1} for consideration of newly
  formed, large-grain dust in the expanding, outer ejecta as a possible
  scatterer; note that pre-existing dust in the nearly stationary surrounding
  CSM is ruled out as a cause due to the observed redshift of the polarized
  flux.}

To test our basic model and draw quantitative conclusions, we performed
time-dependent, non-local thermodynamic equilibrium radiative transfer
simulations of SN ejecta with {\sc CMFGEN} coupled to an upgraded version of
the 2D polarized radiative transfer code {\sc LONG\_POL} \citep[][and
  references therein]{Dessart21a}.  The simulations were carried out in the
manner described by \citet{Dessart21b}, but with model x3p0ext4 selected from
\citet{Hillier19} for the progenitor and ejecta --- a red supergiant (RSG)
progenitor of initial mass ${\rm M}_{\rm init} = 15\ {\rm M}_\odot$ (consistent
with the inferred progenitor of SN~2013ej; \citealt{Dhungana16}) demonstrated
by \citet{Hillier19} to best replicate the photometric and spectroscopic
properties of SN~2013ej.  We aged the explosion to the epoch of our final
observations, and then inserted a region of enhanced free electron density
(i.e., density enhancement factor $N_{\rm e,fac}$ relative to the surrounding
free-electron density) in the outer ejecta to serve as the scattering source.

\begin{figure}[!htbp]
\centering
\rotatebox{90}{
 \scalebox{1.6}{
\includegraphics[width=3.4in]{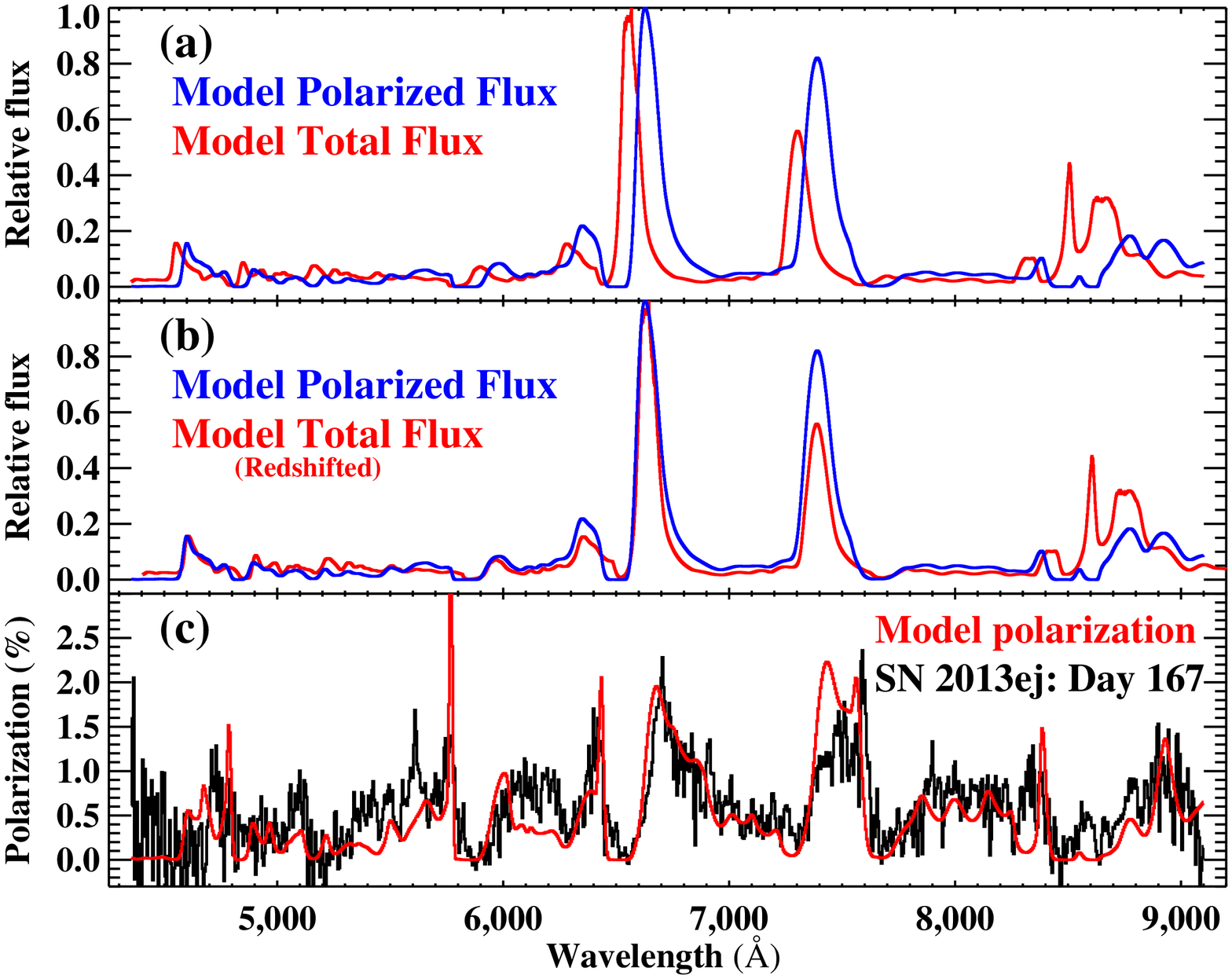} }}
 \caption{Comparing models to observations.  Results from a 2D
   polarized radiative transfer simulation described in the text and designed
   according to the basic physical setup depicted in Figure~\ref{plot309f2} and
   detailed by \citet{Dessart21b}.  The scattering source in this example is
   characterized by $v_{\rm rec} = 4,000$ \kms, $v_{\rm spread} = 400$ \kms,
   $\alpha_{\rm LOS} = 90^\circ$, $2\beta = 30^\circ$, and $N_{\rm e,fac} =
   100$. (a, b) Total and polarized flux resulting from the simulation,
   presented in a manner identical to that shown for the data of SN~2013ej in
   Figure~\ref{plot307f7a}, where we have again normalized all spectra to 1.0
   at the location of the H$\alpha$ peak.  (c) Direct comparison between the
   model's predicted polarization (red, smooth curve) and that actually
   observed for SN~2013ej on Day 167 (black, noisy data). \label{plot306j14}}
   
\end{figure}

Following \citet{Dessart21b}, we ran multiple simulations, allowing $N_{\rm
  e,fac}$, $v_{\rm rec}$, {\bf $\alpha_{\rm LOS}$,} and the region's opening
angle (defined as $2\beta$, where $\beta$ is the half-opening angle) and spread
in radial velocity ($v_{\rm spread}$) to vary and tested each realization for
conformity with observational constraints (Appendix~\ref{appendixc2}).
Redshifted, polarized fluxes showing a clean reflection of the total flux were
routinely obtained in our simulations, provided the scattering region had
sufficient optical depth and was placed well outside of the primary region of
spectrum formation.  Figure~\ref{plot306j14} presents results from one such
simulation, together with a direct comparison with the observed polarization
for SN~2013ej on day 167.  The overall agreement between our models and the
observations in terms of both the specific character and the level of the
observed polarization demonstrate that our basic {\it ansatz} is indeed capable
of replicating the essential observations of SN~2013ej.

Further comparisons between the models and observations permit quantitative
constraints on the location and distribution of the scatterer (or, scatterers)
to be drawn.  First, we note again that the {\it entire} emission profiles,
including asymmetries, appear well replicated in polarized light
(Figures~\ref{plot307f7a} and \ref{plot304f3}).  From the velocity of the blue
edges of the emission profiles relative to line center, we are thus able to
conservatively set the recession velocity of the inner parts of the scatterer,
which typically dominate the redshift imparted to the polarized flux
\citep{Dessart21b}, at $\sim 4,000$ \kms.  Since $v_{\rm shift} \approx
4,000$\ \kms\ for the observations of SN~2013ej, this effectively limits
$\alpha_{\rm LOS}$ to values of $\alpha_{\rm LOS} \lesssim 90^\circ$
(Figure~\ref{plot309f2}); that is, the dominant scattering source responsible
for the polarization must be located close to the plane of the sky or in the
approaching hemisphere of the remnant.

Second, the lack of broadening of the polarized flux profiles of lines relative
to the profiles seen in the total flux (Figure~\ref{plot304f3}) constrains the
opening angle of the scatterer to values of $2\beta \lesssim 40^\circ$, since a
less confined scatterer unacceptably broadened the polarized line profile in
our models (Appendix~\ref{appendixc2}).  This also restricts any ``bi-polar''
(or otherwise multiply distributed) arrangement to the scatterer(s) to
orientations within $\sim 10^\circ$ of the plane of the sky and with nearly
identical values of $v_{\rm rec}$ (see Appendix~\ref{appendixc2}).  To
summarize: Comparing our models with the observed data constrains a single,
asymmetrically distributed scatterer to have $\alpha_{\rm LOS} \lesssim
90^\circ$, $v_{\rm rec} > 4,000$ \kms, and $2\beta < 40^\circ$; if multiple
scatterers are responsible for the polarization, they must all be contained
close to the plane defined by $\alpha_{\rm LOS} \approx 90^\circ$ (the plane of
the sky) and at nearly the same recession velocities.

Establishing the existence of a high-velocity, geometrically confined scatterer
does not establish its cause.  As discussed in Section~\ref{sec1}, a popular
scenario capable of producing enhanced free-electron density in the expanding
ejecta of CC SNe is decay heating from radioactive nickel, created and ejected
during the explosion.  To investigate this possibility, we studied the effects
of a concentration of $^{56}{\rm Ni}$ on the ionization structure of our
specific model at nebular times.  Consistent with the results of
\citet{Dessart21b}, we found that even a relatively small amount ($M_{\rm Ni}
\approx 0.004\ {\rm M}_\odot$), placed in a very confined clump at sufficient
velocity ($v_{\rm rec} \gtrsim 4,000$\ \kms) produced ionization and, hence,
electron density enhancements sufficient to generate the polarization levels
observed (e.g., Figure~\ref{plot306j14}).

The enhanced electron density in our simulations typically extended $\gtrsim
500$ \kms\ beyond the specific location of the nickel clump, into both the
denser, more slowly moving ejecta and the less-dense, more quickly moving
ejecta.  Since it is the recession speed of the inner edge of the enhanced
electron density that is typically responsible for the redshift imparted to the
polarized flux \citep{Dessart21b}, this locates the putative $^{56}$Ni in
SN~2013ej at a minimum recession velocity of $\gtrsim 4,500$ \kms.

\section{Conclusions and Discussion}
\label{sec4}

In this work, we show direct evidence from nebular-phase spectropolarimetry
that a confined, asymmetrically distributed, high-velocity scattering source
exists in the ejecta of the Type II-P/L SN~2013ej.  We find the most plausible
physical origin to be free electrons liberated through ionization caused by the
decay of radioactive $^{56}$Ni, synthesized by and launched as part of the
explosion with final radial velocity exceeding $4,500$ \kms.

This specific finding for SN~2013ej, an otherwise fairly typical CC SN arising
from a progenitor star with a massive hydrogen envelope intact
\citep{Dhungana16,Hillier19}, must be viewed in light of the predictions of
hydrodynamical simulations of neutrino-driven SN explosions.  Calculations in
such stars that incorporate the development of Rayleigh-Taylor instabilities
\citep{Chevalier76}, when carried out assuming an initially spherical SN shock,
have proven capable of obtaining maximum $^{56}{\rm Ni}$ velocities of $\sim
2,000$\ \kms\ \citep{Herant91}.  More recently, such models that treat
consistently the development of explosion asymmetries --- produced by
convective instabilities behind the stalled shock and the standing accretion
shock instability \citep{Blondin03} --- have achieved maximum $^{56}{\rm Ni}$
velocities of between $3,700 - 4,400$\ \kms\ in some instances
\citep{Wongwathanarat15}, with speeds approaching $\sim 5,000$\ \kms\ claimed
in only the most extreme cases \citep{Orlando21}.

In this context, it may also be useful to consider studies of the much more
energetic SNe associated with long-duration $\gamma$-ray bursts (GRBs), which
arise from fast-rotating, higher-mass progenitors (${\rm M_{\rm initial}}
\gtrsim 25{\rm\ M}_\odot$; \citealt{Woosley06}) whose envelopes have been
completely stripped of their hydrogen and helium layers prior to explosion.
Here, detection of extremely high-velocity ($v \gtrsim 100,000$ \kms), Ni-rich
ejections of material \citep{Izzo19} implicate a grossly non-spherical
explosion in which the explosion mechanism --- typically envisaged as requiring
a ``central engine'' \citep{Sobacchi17} --- drives material well beyond the
denuded core of the progenitor star.  Similarly high-velocity material is also
inferred from the characteristics of very early time spectra of some
``stripped-envelope'' CC SNe that are not associated with GRBs (i.e., SNe~Ib/c,
IIb; \citealt{Izzo20}).  The confirmation that a high-velocity scatterer exists
in the thinning ejecta of SN~2013ej may bolster the contention (see
\citealt{Ono20}, and references therein) that a related mechanism is at play in
more typical SNe~II that arise from lower-mass progenitors \citep{Smartt15},
with unequivocal polarization signatures only coming to dominate the signal
once the thick, initially opaque hydrogen envelope has turned sufficiently
transparent.  Additional, high SNR nebular-phase spectropolarimetry of nearby
Type II SNe are certainly warranted to further test this conjecture.

\begin{acknowledgments}
We thank an anonymous referee for helpful comments that improved the
manuscript.  This work was supported by the ``Programme National de Physique
Stellaire'' of CNRS/INSU co-funded by CEA and CNES.  D.J.H. thanks NASA for
partial support through the astrophysical theory grant
80NSSC20K0524. D.C.L. acknowledges support from NSF grants AST-1009571,
AST-1210311, and AST-2010001, under which part of this research was carried
out.  J.L.H. acknowledges support from NSF grants AST-1210372 and AST-2009996.
G.P. acknowledges support by the Ministry of Economy, Development, and
Tourism's Millennium Science Initiative through grant ICN12\_009, awarded to
The Millennium Institute of Astrophysics, MAS.  L.D. is grateful for the access
to the HPC resources of CINES under the allocation 2018 -- A0050410554, 2019 --
A0070410554, and 2020 -- A0090410554 made by GENCI, France.  Based on
observations collected at the European Southern Observatory under ESO programme
091.D-0401(A).  Additional observations reported here were obtained at the MMT
Observatory, a joint facility of the University of Arizona and the Smithsonian
Institution.
\end{acknowledgments}

\vspace{5mm}
\facilities{MMT(SPOL), VLT:Antu(UT1, FORS2)}
\software{{\sc CMFGEN}, {\sc LONG\_POL} \citep[][and
  references therein]{Dessart21a}}

\appendix

\section{Data Acquisition, Reduction, and Correction for ISP}
\label{appendixa}
\renewcommand{\thefigure}{A\arabic{figure}}
\renewcommand{\thetable}{A\arabic{table}}
\renewcommand{\theequation}{A\arabic{equation}}
\setcounter{figure}{0}
\setcounter{table}{0}
\setcounter{equation}{0}
\subsection{Observations and Data Reduction}
\label{appendixa1}
\begin{figure}[!htbp]
\centering
\rotatebox{90}{
 \scalebox{1.65}{
 \includegraphics[width=3.4in]{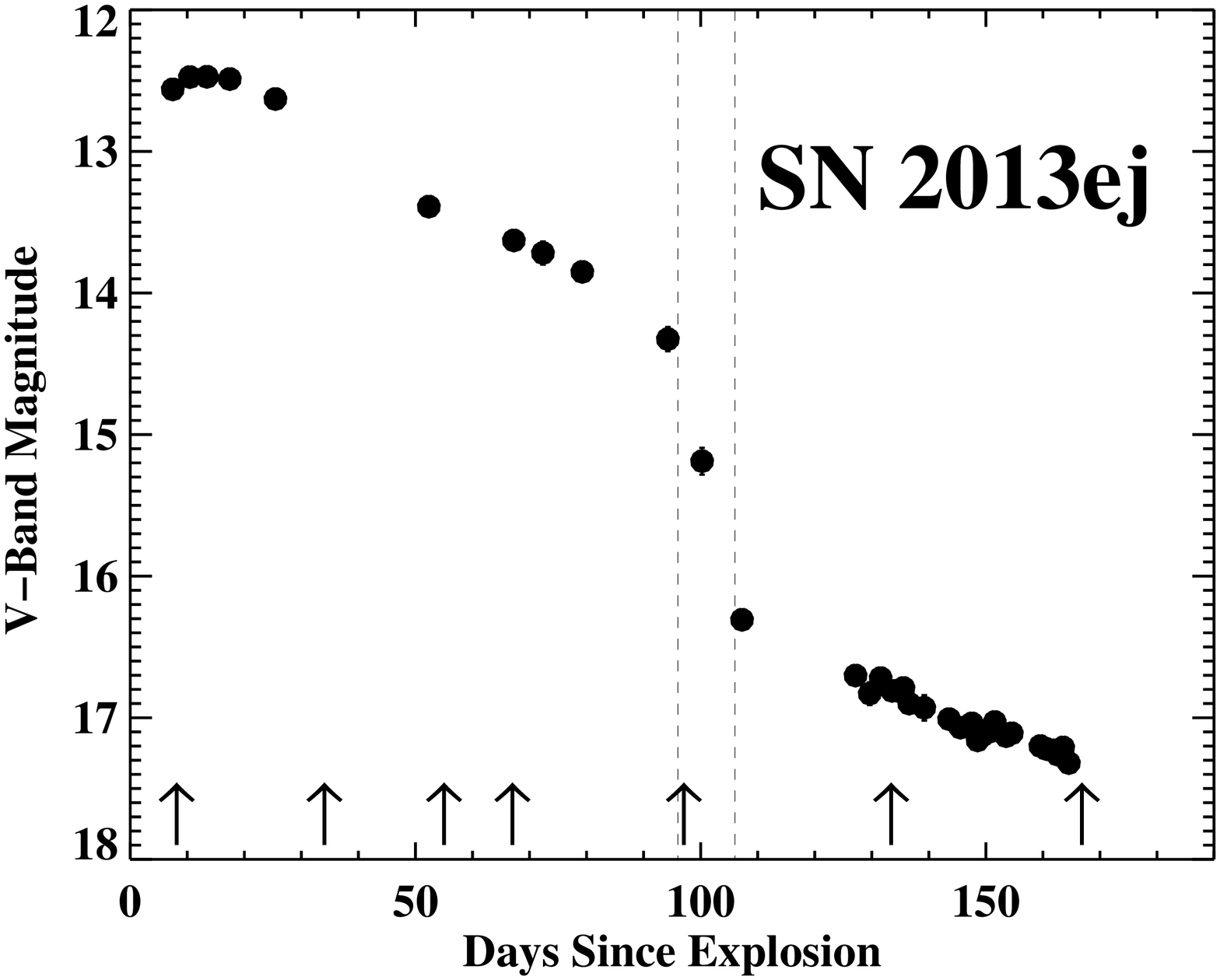} }}
 \caption{{\it V}-band light curve of SN~2013ej (filled circles;
   magnitudes are in the Vega-system) with the timings of the seven
   epochs of spectropolarimetry (Table~\ref{tab:1}) indicated by vertical
   arrows, all relative to the date of explosion estimated by
   \citet{Valenti14}.  The photometric data are taken from the Mount Laguna
   Supernova Survey (MOLASUS; \citealt{Khandrika16}) through day 128, and
   supplemented with data from \citet{Huang15} thereafter. Dashed, vertical
   lines denote the precipitous drop in luminosity that occurs during the
   transition period between the opaque ``photospheric'' phase (before $\sim$
   day 96 for SN~2013ej) and the increasingly transparent radioactive tail
   phase \citep[after $\sim$ day 106;][]{Yuan16}. Note that the $V$-band
   decline of $\sim 0.87$ mag \citep{Bose15} between peak and day 50 after
   explosion places SN~2013ej close to the boundary established by
   \citet{Faran14b} that separates ``Type II-L'' from ``Type II-P''; indeed,
   most authors consider SN~2013ej to be an intermediate, or ``transitional'',
   object \citep{Bose15,Dhungana16,Morozova18}, one of a growing number that
   straddle the defining characteristics of the Type II-P/L
   categories.\label{plot103b}}
\end{figure}

Our complete observational program gathered seven epochs of spectropolarimetry
of SN~2013ej.  All observations were made at the VLT with FORS2 mounted
at the Cassegrain focus of the Antu (UT1) telescope.  The data sample days 8 --
172 following the estimated date of explosion.  Details of the observational
epochs, instrumental setup, exposure times, and observational conditions are
given in Table~\ref{tab:1}.  Note that due to observational constraints
(maximum visibility of SN~2013ej of $\sim 1$ hour by the final observations, as
it set shortly after sunset), the two nebular-phase epochs (i.e., epochs 6 and
7) that are the focus of this study incorporate data taken on multiple nights.
The timing of all of the spectropolarimetric epochs relative to the photometric
light curve of SN~2013ej is shown in Figure~\ref{plot103b}.

The data reduction, including extraction of the spectra, correction for the
small ($\lesssim 0.1\%$) amount of instrumental polarization known to exist at
the VLT \citep{Cikota17}, and formation of the Stokes parameters and total flux
spectra, was carried out according to the procedures described by
\citet{Dessart21a}.  From each epoch's final observed dataset, we subtracted
the ISP derived in Appendix~\ref{appendixa2}, removed a redshift of 657
\kms\ \citep[][via NED]{Lu93}, and rebinned to $5$~\AA\ bin$^{-1}$ over a
wavelength range of $4300$ \AA\ -- $9100$ \AA.

\begin{deluxetable}{lccccccc}
\tablewidth{480pt}
\tablecaption{ {\bf Journal of Spectropolarimetric Observations of SN~2013ej} }
\tablehead{
\colhead{}  &
\colhead{}  &
\colhead{}  &
\colhead{}  &
\colhead{Exposure\tablenotemark{d}} &
\colhead{Air} &
\colhead{Seeing\tablenotemark{f}} &
\colhead{Slit}\\
\colhead{Epoch}  &
\colhead{Day\tablenotemark{a}}  &
\colhead{UT Date\tablenotemark{b}}  &
\colhead{MJD\tablenotemark{c}}  &
\colhead{(s)} &
\colhead{Mass\tablenotemark{e}} &
\colhead{(arcsec)} &
\colhead{(arcsec) }}
\startdata
1       & 8.442   & 2013-08-01.392 & 56,505.392 & 5,760\tablenotemark{g} (10)  & 1.31--1.40 & 1.0 & 1.0\\
2       & 34.346  & 2013-08-27.296 & 56,531.296 & 3,600 (8)  & 1.31--1.45 & 1.0 & 1.5\\
3       & 55.277  & 2013-09-17.227 & 56,552.227 & 3,680 (5)  & 1.33--1.49 & 1.0 & 1.5\\
4       & 67.307  & 2013-09-29.257 & 56,564.257 & 6,400 (21) & 1.31--1.51 & 0.8 & 1.5\tablenotemark{h}\\
5       & 97.297  & 2013-10-29.247 & 56,594.247 & 7,200 (15) & 1.35--2.34 & 1.2 & 1.0\\
6       & 133.150 & 2013-12-04.100 & 56,630.100 & 9,600 (2)  & 1.31--1.61 & 1.7 & 1.6\\
\nodata & 134.144 & 2013-12-05.094 & 56,631.094 & 8,050 (2)  & 1.31--1.51 & 1.0 & 1.6\\
7       & 163.106 & 2014-01-03.056 & 56,660.056 & 3,600 (1)  & 1.43--1.64 & 1.0 & 1.0\\
\nodata & 164.113 & 2014-01-04.063 & 56,661.063 & 3,600 (2)  & 1.46--1.76 & 1.0 & 1.6\\
\nodata & 172.105 & 2014-01-12.055 & 56,669.055 & 3,600 (2)  & 1.54--1.93 & 1.0 & 1.6\\
 \enddata

\tablecomments{All observations were obtained with VLT-FORS2 using the 300
  line/mm grism (``GRIS\_300V'') along with the ``GG435'' order-sorting filter
  to prevent second-order contamination.  This resulted in a useable spectral
  range of 4300 \AA\ -- 9200 \AA.  Observations made through a 1\farcs0 slit
  delivered a resolution of $\sim12$\ \AA\ ($R \approx 550$ at H$\alpha$),
  whereas those made through 1\farcs5 and 1\farcs6 slits delivered resolutions
  of $\sim17$\ \AA\ and $\sim18$\ \AA\ ($R \approx 350$ at H$\alpha$),
  respectively, derived from the full width at half maximum (FWHM) of night-sky
  lines.}

\tablenotetext{a}{Day since the estimated explosion date of 2013-07-23.95 UT = MJD
  56,496.95 = JD 2,456,497.45 \citep{Valenti14}, taken as the mean time of
  observation of the complete set of data for each epoch (i.e., the midpoint of
  all of the individual exposures). For epochs 6 and 7, for which data were
  obtained over multiple nights, the mean ``days since explosion'' are 133.647
  (epoch 6) and 167.108 (epoch 7).}

\tablenotetext{b}{yyyy-mm-dd.ddd}

\tablenotetext{c}{Modified Julian Date (Julian Date - 2400000.5).}

\tablenotetext{d}{Total exposure time in seconds, with the number of complete
  sets of 4 waveplate positions obtained shown in parenthesis.}

\tablenotetext{e}{Air mass range for each night of observations.}

\tablenotetext{f}{Average value of the FWHM of the spatial profile on the CCD chip for each
  night, rounded to the nearest 0\farcs1.}

\tablenotetext{g}{The combined exposure time of all 10 complete
observational sets is reported.  However, due to observational error on this
  night, 6 of the sets were affected by saturation over parts of the spectral
  range, generally shortward of 6600\ \AA.  We therefore carefully used only
  the unsaturated parts of the spectra when forming the final polarization
  data. Thus, the total exposure time on this date is 5,760 sec for $\lambda >
  6600$\ \AA , but ranges down to 2,160 sec for $\lambda < 5500$\ \AA, with
  intermediate total exposure times for $5500$\ \AA $< \lambda < 6600$\ \AA.}

\tablenotetext{h}{Note that the first two complete sets were obtained through a
  1\farcs0 slit.}

\label{tab:1}

\end{deluxetable}

 To translate the normalized Stokes $q$ and $u$ parameters into estimates of
 polarization, $p$, we computed the polarization according to each of the four
 definitions given by \citet{Leonard3}: ``Traditional Polarization'' ($p_{\rm
   trad}$), ``Debiased Polarization'' ($p_{\rm deb}$), ``Optimal Polarization''
 ($p_{\rm opt}$), and ``rotated Stokes parameter'' (RSP).  As expected for data
 with a strong, non-zero polarization signal, all four calculations yielded
 very similar results when applied to the observed data.  Following
 \citet{Leonard3}, we calculated the RSP by first smoothing the polarization
 angle curve and then rotating the Stokes parameters through the
 wavelength-dependent angle defined by the smoothed curve. This procedure is
 designed to place all of the polarization signal in the RSP, which makes it
 effectively equivalent to $p_{\rm trad}$ in regions where the SNR and
 polarization are high, but with improved noise characteristics in spectral
 regions with low polarization levels, which exist in our data following
 removal of the ISP (see Figure~\ref{plot306j14}c).  For both presentation and
 computational purposes, we therefore adopted the RSP as our best estimate of
 $p$, including for our calculations of the polarized flux.  That is, the
 polarized flux spectra presented and analyzed in this work were produced by
 multiplying the RSP by the total flux: ${\rm RSP}(\lambda) \times f(\lambda)$.
 We note that using any of the other estimates for $p$ had minor impact on the
 resulting polarized flux, except in spectral regions with very low
 polarization, in which the RSP is much better behaved and, thus, preferred.

\subsection{Interstellar Polarization}
\label{appendixa2}

\begin{figure}[!htbp]
\centering
\rotatebox{90}{
 \scalebox{1.7}{
 \includegraphics[width=3.4in]{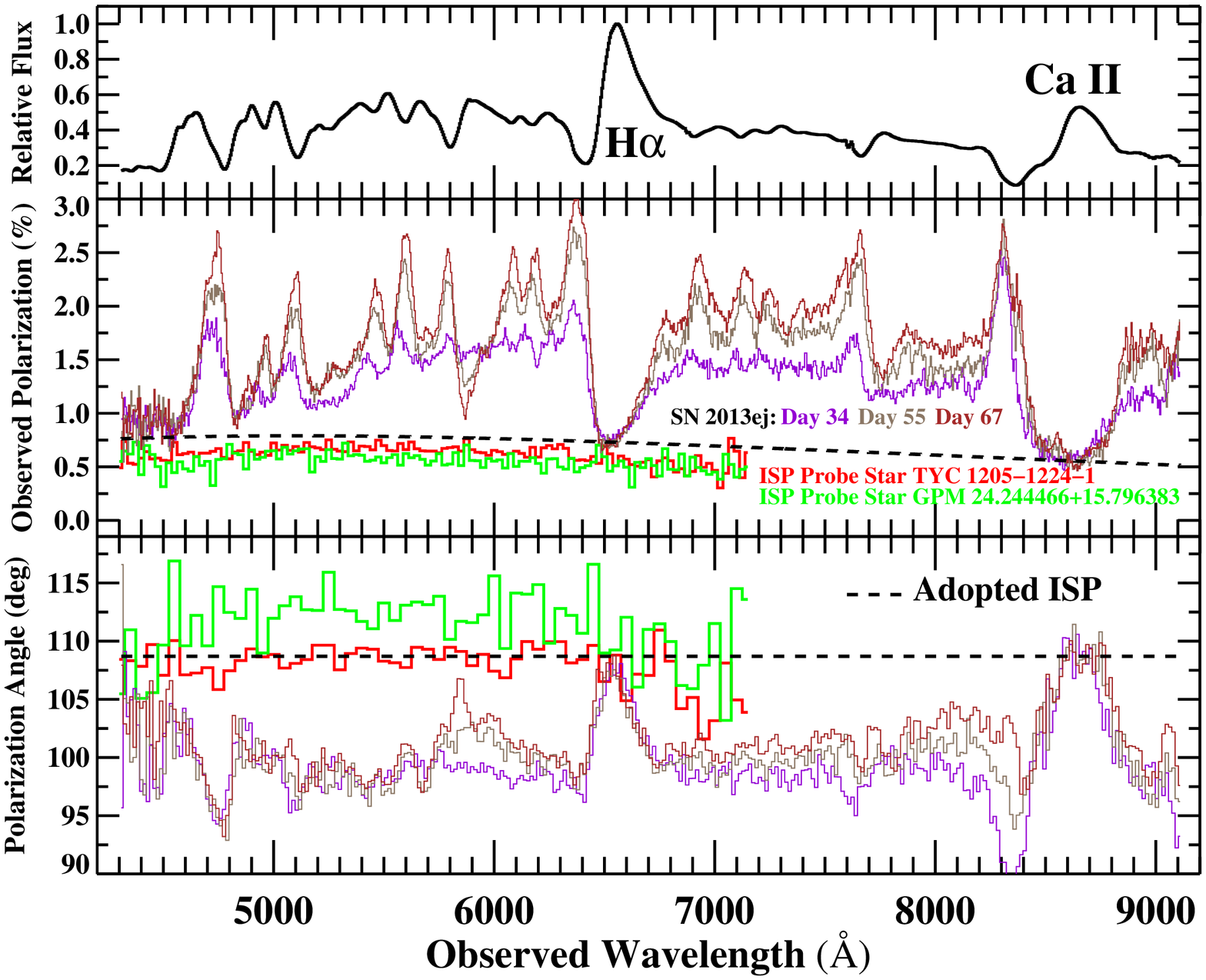} }}
 \caption{Determining the ISP to SN~2013ej.  {\it Top}: Total flux spectrum of
   SN~2013ej taken on day 55 after explosion, with prominent emission lines
   labeled.  {\it Middle}: Observed polarization level (thin lines) for three
   mid-photospheric-phase epochs, color coded by day since explosion
   (Table~\ref{tab:1}).  {\it Bottom}: Same as for {\it middle}, but for the
   observed polarization angle. Also shown in both the middle and bottom panels
   are the observed polarizations and polarization angles of two Galactic ``ISP
   probe'' stars TYC~1205-1224-1 (thick, red lines) and GPM~24.244466+15.796383
   (thick, green lines); note that these data do not extend redward of
   $7150$\ \AA.  The total ISP derived in Appendix~\ref{appendixa2} is
   indicated by the thick, black, dashed line in both panels. \label{plot218g}}
\end{figure}

To derive the ISP for SN~2013ej, we followed the approach detailed by
\citet{Dessart21a}, in which strong, unblended emission features (e.g., regions
near the peaks of H$\alpha$ and the \ion{Ca}{2} IR triplet) are assumed to
completely depolarize all SN light during the optically thick photospheric
phase, leaving only the ISP.  For multi-epoch data, this should present as
fixed, unchanging levels of polarization (and, polarization angle) with time in
these spectral regions.  Figure~\ref{plot218g} displays our data taken during
the middle of the photospheric phase of SN~2013ej, on days 34, 55, and 67 after
explosion (Table~\ref{tab:1}).  Sharp, consistent depolarizations are indeed
seen in the spectral regions near the peaks of the H$\alpha$ and \ion{Ca}{2} IR
triplet lines in the total flux spectrum (i.e., near $6540$\ \AA\ and
$8620$\ \AA, respectively).  What is more, while the overall polarization level
(and, to a lesser extent, the polarization angle) shows substantial evolution
over the 33 days covered by these observations, these specific regions do not,
always returning to the same polarization level and angle, regardless of epoch.
Such constancy is fully consistent with expectations that the total, unchanging
ISP is indicated at these wavelengths during these epochs.

For an SN with minimal extra-Galactic reddening, one way to test any ISP
estimate is to observe Galactic ``ISP probe'' stars \citep{Tran95}: Distant,
intrinsically unpolarized stars in the Milky Way that lie near to the LOS of
the SN and are sufficiently far away to fully probe the ISP produced by
Galactic dust.  SN~2013ej is an excellent candidate for such a test, since it
lacks any detectable interstellar \ion{Na}{1} D $\lambda\lambda 5890, 5896$
\AA\ absorption at the redshift of its host galaxy (NGC~628) in high-resolution
spectra, which implies negligible host-galaxy extinction
\citep{Valenti14,Bose15} and, by logical extension, host-galaxy ISP.  We thus
obtained spectropolarimetry of two distant Galactic stars using the 6.5 meter
Multiple Mirror Telescope (MMT) on Mt. Hopkins on 2016 July 8 (UT), as part of
the Supernova Spectropolarimetry (SNSPOL; \citealt{Hoffman17}) Project.  The
data were acquired and reduced as described by \citet{Mauerhan14}.  The first
``probe'' star, GPM 24.244466+15.796383, is separated by only $3\farcm4$ from
SN~2013ej in the sky and is estimated by Gaia parallax \citep{Bailerjones18} to
be $700 \pm 30$ pc away.  The second star, TYC 1205-1224-1 is separated by
$53\farcm3$, and is estimated by Gaia parallax to be $319 \pm 5$ pc away.  Both
stars satisfy the criterion of \citet{Tran95} that they be located at least 150
pc above or below the Galactic plane in order to fully sample the Galactic ISP
along the LOS. (The Galactic latitude of SN~2013ej, $b = -45.7^\circ$,
necessitates a distance of at least $210$ pc.)

We overplot the ``probe'' star spectropolarimetry data in the bottom panel of
Figure~\ref{plot218g}, and find them to show remarkable consistency with expectations
based on the depolarized spectral line analysis: The measured polarization
levels and polarization angles of both stars are very close to those measured
in the H$\alpha$ peak of SN~2013ej during the photospheric epochs.  (The
limited spectral range precludes such a direct comparison at the location of
\ion{Ca}{2}.)  Small differences may presumably be attributed to either patchy
Galactic ISP or a small amount of host galaxy ISP.

With the overall level and angle of the ISP thus convincingly constrained, we
proceeded to derive a total ISP to be subtracted from our data as follows.
First, we estimated the observed polarization near the peaks of both H$\alpha$
(6520\ \AA\ -- 6560\ \AA) and \ion{Ca}{2} (8600\ \AA\ -- 8640\ \AA) by taking
the average values obtained from all three mid-photospheric epochs. Next, we
assumed the empirically derived functional form for ISP given by
\citet{Serkowski73}:

\begin{equation}
p_{\rm ISP} = p_{\rm max,ISP} \exp[-K_{\rm ISP}\ln^2(\lambda_{\rm
    max,ISP}/\lambda)]
\label{eqn:1}
\end{equation}
\noindent where $p_{\rm ISP}$ is the percent polarization at a wavelength
$\lambda$, $p_{\rm max,ISP}$ is the maximum level of ISP that occurs at
$\lambda_{\rm max,ISP}$ , and $K_{\rm ISP}$ is a constant that determines the
full width of the linear polarization curve.  We arrived at the level and
wavelength dependence for the ISP across our full spectral range then by
seeking the set of values of $\lambda_{\rm max,ISP},\ p_{\rm max,ISP}$, and
$K_{\rm ISP}$ that produced the best-fitting ``Serkowski curve'' (i.e.,
Equation~\ref{eqn:1}) to our two ISP points derived from the H$\alpha$ and
\ion{Ca}{2} regions.  Note that rather than adopting an empirical relation
between $K_{\rm ISP}$ and $\lambda_{\rm max,ISP}$ \citep[e.g.,][]{Cikota17}, we
chose instead to allow $K_{\rm ISP}$ to range between 0.6 and 1.5, based on the
ranges seen in recent studies \citep[see][and references therein]{Cikota18};
this also provided a much better fit to our data points.  Finally, guided by
the wavelengths of peak polarization observed for the Galactic probe stars
(Figure~\ref{plot218g}), we restricted $\lambda_{\rm max,ISP}$ to lie between
$5,000 - 5,500$\ \AA .

Our minimization resulted in best-fit values of $p_{\rm max,ISP} = 0.79\%$,
$\lambda_{\rm max,ISP} = 5100$\ \AA, and $K_{\rm ISP} = 1.28$; we set the
polarization angle to be $\theta_{\rm ISP} = 108.7^\circ$, the simple average
of the values obtained over the H$\alpha$ and \ion{Ca}{2} regions in our three
epochs.  This adopted ISP is overplotted in Figure~\ref{plot218g}, and
represents the ISP that was then subtracted from all observed data to produce
the intrinsic SN polarization presented and analyzed by this work.  The close
agreement between the ISP derived from the polarization observed for SN~2013ej
itself with that measured from the probe stars, coupled with the evidence for
minimal host galaxy ISP, provides confidence in our chosen ISP and the
technique used to derive it; small, but justifiable, changes to it were tested
and found to have negligible impact on the quantitative results of this work.

We note that our adopted ISP is similar to the ISP value derived by
\citet{Nagao21} in their spectropolarimetric study of SN~2013ej, but differs
somewhat from the value chosen by \citet{Mauerhan17}.  As pointed out by
\citet{Nagao21}, the discrepancy may not be that surprising given that
\citet{Mauerhan17} adopted the ISP derived towards a different SN (SN~2002ap)
in the same host galaxy, for which significant host galaxy ISP is suspected
\citep[see][and references therein]{Nagao21}.

\section{Data Analysis}
\label{appendixb}
\renewcommand{\thefigure}{B\arabic{figure}}
\renewcommand{\thetable}{B\arabic{table}}
\setcounter{figure}{0}
\setcounter{table}{0}
\subsection{Deriving the Velocity Shift of the Polarized Flux}
\label{appendixb1}

To derive the (relativistic) velocity offset between the total and polarized
flux spectra, we ran IRAF's FXCOR task (``Release Mar93'') with the polarized
flux spectrum as the ``object'' and the total flux spectrum as the
``template''.  We turned on continuum subtraction for both the object and
template (not doing so altered the calculated shift by less than 20 \kms\ in
all cases), and ran the correlation over the entire spectral range (4300 --
9100 \AA).  Following \citet{Filippenko99}, we divided the uncertainty
reported by the task by a factor of 2.77, to bring the uncertainties in line
with those calculated using equation (20) of \citet{Tonry79}.

\begin{figure}[!htbp]
\centering
\rotatebox{90}{
 \scalebox{1.6}{
 \includegraphics[width=3.4in]{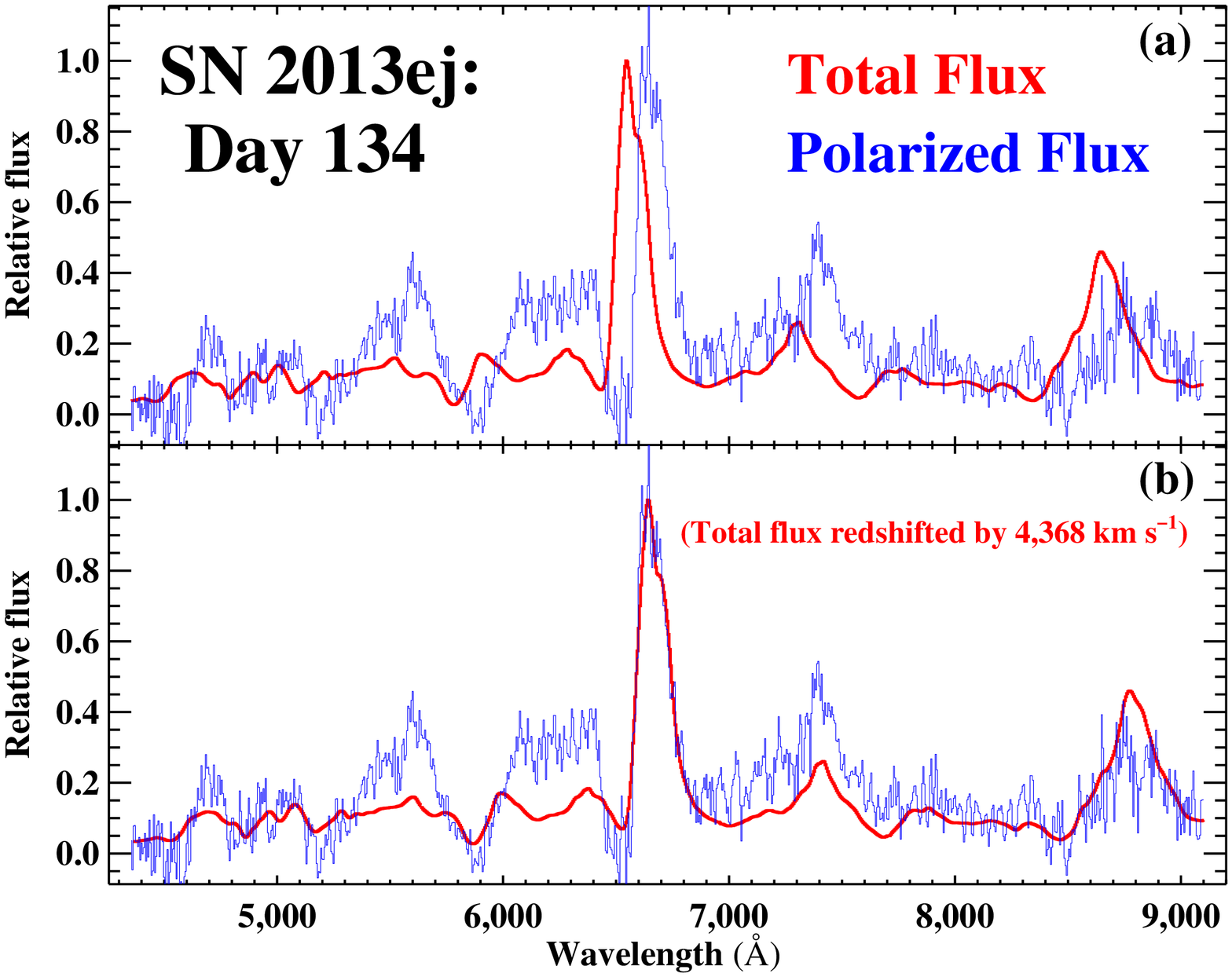} }}
 \caption{Comparing the total and polarized flux spectra of
   SN~2013ej on day 134, displayed as in Figure~\ref{plot307f7a} for the day
   167 data.  To aid comparison of features, the total flux spectrum has been
   scaled by a factor of 0.0035, chosen to produce the best fit to the
   H$\alpha$ spectral region; the total flux spectrum has been artificially
   redshifted by $4,368$ \kms\ in (b).  \label{plot307f6a}}
\end{figure}

Table~\ref{tab:2} presents the results of our cross-correlations for all seven
epochs of data.  In general, normalized cross-correlation peak values above 0.8
(80\% correlation) are considered ``excellent'', whereas those below $\sim 0.5$
are not considered to be ``good'' \citep{Alpaslan09}.  Significant correlation
between the total and polarized flux exists only during the two nebular-phase
epochs, on days 134 and 167.  For completeness, we show in
Figure~\ref{plot307f6a} the data for the day~134 epoch in a manner identical to
that shown in Figure~\ref{plot307f7a} for the day~167 data.

From the lack of correlation prior to the nebular phase, our proposed mechanism
for the SN polarization --- reflection off of an asymmetrically distributed,
externally located, high-velocity scatterer --- appears to dominate the
polarization signal only at nebular times.  We have thus focused our analysis
here squarely on the nebular data.  We note that the photospheric-phase data,
some of which are displayed in Figure~\ref{plot218g} where they are utilized to
derive the ISP, are the same data analyzed by \citet{Nagao21} in their study of
the continuum polarization of SN~2013ej during the photospheric phase.  The
primary characteristics of the nebular-phase polarization data that are
uniquely analyzed by our study --- i.e., redshifted, polarized flux observed
across all strong line features --- are not present at the earlier times
investigated by either \citet{Nagao21} or \citet{Mauerhan17}, and thus demand a
distinctly different analysis.  We leave an evaluation of the possible
contribution that a high-velocity scatterer could have had on the
photospheric-phase spectropolarimetry to a future work (D. Leonard et al., in
preparation).

\begin{deluxetable}{llll}
\tablewidth{0pt}

\tablecaption{ {\bf Measured Correlation Between Total Flux and Polarized Flux} }
\tablehead{
\colhead{}  &
\colhead{}  &
\colhead{Normalized Cross-} &
\colhead{Peak}\\
\colhead{Epoch\tablenotemark{a}}  &
\colhead{Day}  &
\colhead{Correlation Peak} &
\colhead{Significance\tablenotemark{b}}}

\startdata
1 & 8  &  0.341 &  1.6   \\
2 & 34 &  0.280 &  1.7   \\
3 & 55 &  0.309 &  1.5   \\
4 & 67 &  0.298 &  1.3   \\
5 & 97 &  0.336 &  1.5   \\
6 & 134 & 0.843 &  7.6   \\
7 & 167 & 0.915 &  14.6  \\
 \enddata

\tablenotetext{a}{As defined by Table~\ref{tab:1}.}
\tablenotetext{b}{As quantified by the \citet{Tonry79} {\it R} value.}

\label{tab:2}

\end{deluxetable}

\subsection{Constraints on the Broadening of Line Features in the Polarized
  Flux} 
\label{appendixb2}

For comparison with models (see Appendix~\ref{appendixc}), a critical
observational constraint is the degree to which unblended emission-line
features appearing in the polarized flux are {\it not} broadened relative to
the same features in the total flux \citep{Dessart21b}.  Casual examination of
Figure~\ref{plot307f7a} and Figure~\ref{plot307f6a} in the Appendix reveals no
such obvious broadening.  Further, the close agreement between the total and
polarized flux line-profile morphologies (Figure \ref{plot304f3}) convincingly
argues in favor of the entire flux profile being reflected rather than a
narrower component being broadened by a (finely tuned) spatially extended
scatterer.

Assuming this to be the case, we set quantitative limits on any broadening by
carrying out the following analysis on the H$\alpha$ profile from the day 167
data.  First, we removed the observed velocity shift (i.e., $v_{\rm shift} =
3,946$\ \kms) from the polarized flux.  Next, we rebinned both the total and
polarized fluxes to the approximate spectral resolution (Table~\ref{tab:1}),
and then converted the wavelength scale to a (relativistic) velocity scale with
respect to the rest wavelength of H$\alpha$ (i.e., $6562.8$\ \AA).  We then
sought the scaling to apply to the total flux to provide the best match between
it and the polarized flux (accounting for uncertainties) over the velocity
range of $\pm 5,000$ \kms.  Minimizing the $\chi^2$ of the fit yielded a
scaling of $0.0041$, which we then applied to the total flux spectrum.  This
scaling produced a reduced $\chi^2$ of $\chi^2_{\rm red} = 1.027$ over this
velocity range which, for 15 degrees of freedom (i.e., 16 data points and 1
free parameter), implies disagreement between the total and polarized fluxes at
the $0.80\sigma$ level for a one-dimensional Gaussian (see discussion by
\citealt{Leonard18}) --- a convincing level of agreement.

We then applied ``broadening factors'' to the observed polarized flux by
multiplying the velocity vector by factors ranging from 0.80 to 1.50, sampling
by hundredths.  For each realization, we minimized $\chi^2_{\rm red}$, keeping
the scaling as a free parameter for each fit.  

Next, we sought any evidence that the polarized flux was {\it already}
broadened relative to the total flux, which would logically manifest as better
fits for broadening factors of less than unity; no such evidence was found.  On
the other hand, the fits did slightly improve for broadening factors greater
than unity, up through a factor of 1.07 (i.e., a 7\% broadening), for which
$\chi^2_{\rm red} = 0.72$.  However, given that measurements of $\chi^2_{\rm
  red}$ have an inherent uncertainty of $\sim \sqrt{2/N}$
\citep{Andrae10}, where {\it N} is the number of data points ($N = 16$ in
this case, resulting in an uncertainty in $\chi^2_{\rm red}$ of $\approx
0.35$), the improvement is only significant at the $1\sigma$ level.

To set an upper bound on the possible broadening of the polarized flux we
searched for the broadening factor that would yield a $3\sigma$ discrepancy
between the total flux and the polarized flux for data characterized by our
uncertainties; this occurred at a broadening factor of 1.24.  We therefore take
a $24\%$ broadening as a conservative upper limit on the possible broadening
factor of the polarized flux relative to the total flux that could be present
in our data.  (Note that this limit gets tighter the higher the SNR of the
data: For data with twice the SNR as ours, the limit on the possible broadening
factor is only 1.12; for nine times the SNR it is only 1.06.)

\section{Modeling and Interpretation}
\label{appendixc}
\renewcommand{\thefigure}{C\arabic{figure}}
\renewcommand{\thetable}{C\arabic{table}}
\setcounter{figure}{0}
\setcounter{table}{0}
\subsection{Theoretical Modeling and Consideration of Dust}
\label{appendixc1}

 The numerical simulations presented by \citet{Dessart21b} establish that a
 clump of $^{56}{\rm Ni}$ inserted into realistic SN ejecta is capable of
 generating high enough levels of ionization at nebular times to provide
 sufficient electron scattering to generate overall polarization at the $\sim
 0.5\%$ level seen in our data for SN~2013ej.  Here we employed the same code
 in an effort to: (1) verify the fundamental results of \citet{Dessart21b}
 when using a progenitor and ejecta model specific to SN~2013ej; and (2) set
 quantitative constraints on the distribution of the high-velocity scatterer
 inferred to exist in its ejecta.

We began with the set of model RSG progenitors presented by \citet{Hillier19},
and selected model x3p0ext4 for our direct comparisons since it was shown in
that work to most faithfully reproduce the specific light curves and spectral
evolution of SN~2013ej itself.  To test whether a single, rapidly receding
scatterer consisting of enhanced free-electron density could reproduce the
polarization observed in the nebular-phase data of SN~2013ej we first aged this
model to day 173 and then ran simulations over a wide range of parameter
values.  Similar to the results of \citet{Dessart21b}, we found the simulations
to consistently produce appropriately redshifted, polarized flux spectra,
provided the scatterer was placed sufficiently outside (i.e., at velocities
$\gtrsim 3,000$ \kms; see Figures~8 and 9 of \citealt{Dessart21b}) of the
primary region of spectrum formation; placing the scatterer at velocities
substantially less than this did not replicate the polarized line flux profile
of, e.g., H$\alpha$, due to the substantial overlap between the line-emitting
and line-scattering regions. We emphasize that our simulations are crafted
specifically to investigate the polarization implications of an asymmetrically
distributed, {\it high-velocity} scattering source.  The effects of any
asymmetry in the `core' emission region --- anticipated to be minor on the
nebular-phase polarization but potentially significant for, e.g., line-profile
morphology --- will be addressed by a future work (L. Dessart et al., in
preparation).

\begin{figure}[!htbp]
\centering
\rotatebox{90}{
 \scalebox{1.6}{
 \includegraphics[width=3.4in]{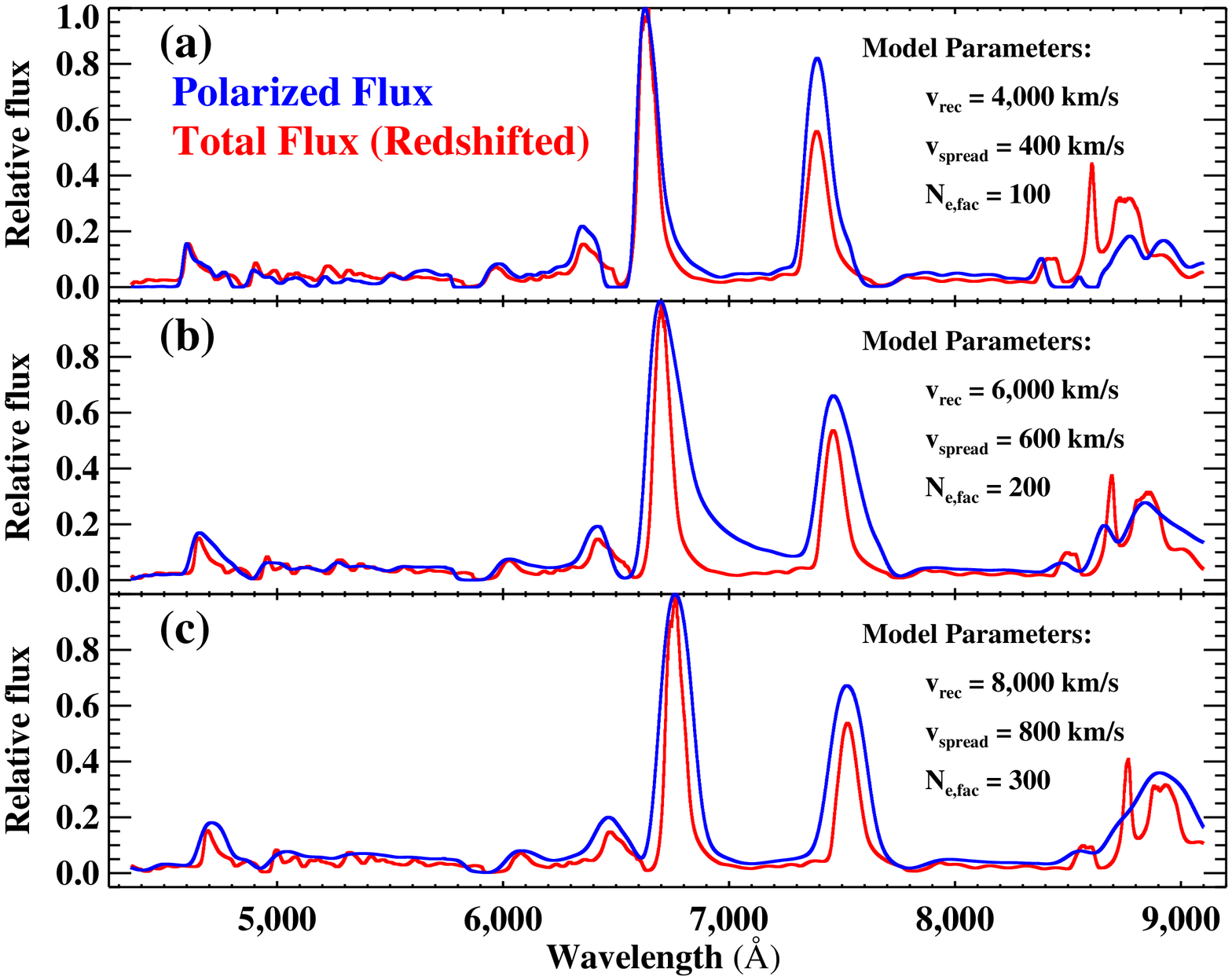} }}
 \caption{Example total and polarized flux spectra drawn from simulations.
   Total flux (red) and polarized flux (blue) for three simulations, with the
   total flux shown in each case artificially redshifted to best match the
   polarized flux.  Each simulation is calculated with the same ejecta
   structure (RSG model x3p0ext4, from \citealt{Hillier19}) and age (173
   days). All ejecta include a concentration of enhanced free-electron density
   inserted at $\alpha_{\rm LOS} = 90^\circ$ (Figure~\ref{plot309f2}) and with
   opening angle $2\beta = 30^\circ$, but with differing recession velocities
   ($v_{\rm rec}$), spreads in radial velocity ($v_{\rm spread}$), and
   free-electron density enhancements ($N_{\rm e,fac}$), as indicated in each
   panel.  All models yield a polarized flux spectrum appropriately redshifted
   with respect to the total flux spectrum, and achieve mean polarization
   levels, calculated over the range 4300 \AA\ $\rightarrow$ 9100 \AA, of the
   same order as that observed in SN~2013ej on day 167: i.e., 0.47\%, 0.58\%,
   and 0.20\% for the models shown in (a), (b), and (c), respectively, compared
   with 0.56\% measured for SN~2013ej.  (Note that a direct comparison between
   SN~2013ej and the polarization generated by the model depicted here in panel
   {\bf a} is shown in Figure~\ref{plot306j14}c.)  There is clearly a complex
   interplay among overall polarization level, breadth of unblended line
   features in polarized flux (e.g., H$\alpha$), and fidelity of line features
   seen in polarized flux (particularly evident in the \ion{Ca}{2} near-IR
   lines).\label{plot306j6d}}
\end{figure}

As an example, one realization from our modeling that was found to reasonably
replicate both the level and the specific features seen in the polarization of
SN~2013ej is shown in Figure~\ref{plot306j14}; Figure~\ref{plot306j6d} shows
further examples that provide a sense of how the resulting polarized flux
varied with changes in the model's input parameters (here modifying $v_{\rm
  rec}$, $v_{\rm spread}$, and $N_{\rm e,fac}$).  Degeneracies certainly exist
among the model parameters.  For instance, models with $\alpha_{\rm LOS} \neq
90^\circ$ can achieve quantitatively similar results to those with $\alpha_{\rm
  LOS} = 90^\circ$ (i.e., the angle that intrinsically generates maximum
polarization and used for the models shown in the figures) with appropriate
adjustments to $v_{\rm rec}$ and $N_{\rm e,fac}$; the model presented in
Figure~\ref{plot306j14} is not a unique, ``best-fit''.  Furthermore, departures
from naive expectation based on the simplified toy model presented by
Figure~\ref{plot309f2} are also clearly apparent in both the simulations as
well as in the data.  For instance, in the model results shown in
Figure~\ref{plot306j6d}, the polarized flux spectra are not perfect
representations of the total flux spectra, deviating notably in the relative
strengths and fidelity of specific line features.  Similar departures are also
seen in the data of SN~2013ej (see Figure~\ref{plot307f7a} and
Figure~\ref{plot307f6a} in this Appendix).  As discussed by \citet{Dessart21b},
such differences may be expected given that the line emitting region ---
particularly for the strongest lines --- is extended out into the ejecta (and
even beyond the region containing the scatterer); it is not the idealized
central source depicted in Figure~\ref{plot309f2}.

As also discussed by \citet{Dessart21b}, many of the qualitative results of
these simulations could plausibly obtain as well with high-velocity dust
serving as the scatterer, since dust is very efficient at scattering, and
polarizing, photons.  However, a main observational constraint demanded by
our observations is wavelength-independence of the scattering source
(Section~\ref{sec3}) which, while a natural consequence of electron scattering,
is not a generic feature of dust scattering unless the grains are much larger
than the wavelengths of light being scattered.  While no specific studies have
yet been carried out that model the wavelength-dependence expected for
large-grain dust scattering at optical wavelengths, investigations at longer
wavelengths find that the diameters of the smallest grains need to be at least
$\sim 3$ times larger than the observed wavelengths to produce ``gray''
scattering \citep[e.g.,][]{Natta07}.  If applied to this situation, it would
require grains of at least $\sim 3\ \mu{\rm m}$ in diameter, with little
tolerance for any significant contribution by smaller-sized grains.

Such dust would need to be created in the high-velocity, outer ejecta of the
SN, most plausibly as the result of interaction between the ejecta and CSM
creating a cold, dense shell within which the dust could conceivably form.
While there is some observational evidence for early dust formation ---
sometimes with large ($\gtrsim 1\ \mu{\rm m}$) inferred grain size --- in the
ejecta of some SNe~II \citep[see][and references therein]{Priestley20a}, the
events typically show signs of interaction with a far denser CSM than that
inferred to exist in SN~2013ej \citep{Bose15,Morozova18}.  Further, the lack of
broadening of the polarized flux features (Appendix~\ref{appendixb2}), as well
as their lack of evolution (Section~\ref{sec3}), would necessitate a single,
persistent dust blob (or two or more such blobs at identical $v_{\rm shift}$),
in contrast with basic expectations of dust formed through interaction with a
more widely distributed CSM.  In contrast, there are clear physical mechanisms
that may produce $^{56}{\rm Ni}$ clumps or ``fingers'', which would naturally
yield isolated geometrical entities.  Thus, although we can not completely rule
out dust as a possibility, we consider it an unlikely agent to explain the
specific scattering characteristics that we see here, and have therefore
quantitatively pursued free electrons as the scatterer at work in SN~2013ej at
these epochs; we postpone to a future study any detailed modeling of dust and
its possible impact on the SN radiation, including its possible effect on line
profile morphology as well \citep[e.g.,][]{Bevan16}.

\subsection{Constraints Derived from the Modeling}
\label{appendixc2}

Although limited by the degeneracies described in the previous section, our
simulations revealed the following three constraining features, which permit us
to draw some conclusions about the scattering source at work in SN~2013ej.

\begin{enumerate}

 \item {\bf The scattering that dominates the polarization signal occurs near
   the inner boundary of the region of enhanced scattering.}  This is likely
   due to the dominating contribution to the scattering from the low-velocity
   portion of the region, where free electron densities are necessarily higher
   for a given value of $N_{\rm e,fac}$.  The amount of the measured difference
   between $v_{\rm rec}$ and $v_{\rm shift}$ was typically $\sim 500 $
   \kms\ (see Figure~\ref{plot306j14}) for free-electron distributions with
   $v_{\rm spread} = 1,000$ \kms, the value typically obtained for ionization
   produced by a confined mass of $^{56}$Ni in our simulations.  This thus
   requires the actual velocity of the putative $^{56}$Ni to be at a somewhat
   larger value than that indicated by the observed velocity shift of the
   polarized flux alone.  Since our observations of SN~2013ej indicate $v_{\rm
     shift} \approx 4,000$ \kms, and the full-profile reflection demands that
   $v_{\rm rec} \gtrsim 4,000$ \kms, this necessarily restricts $\alpha_{\rm
     LOS}$ to values $\lesssim 90^\circ$ (i.e., since $v_{\rm shift} = v_{\rm
     rec}[1 - \cos{\alpha_{\rm LOS}}]$; see Figure~\ref{plot309f2}), and places
   the putative $^{56}$Ni in SN~2013ej at a minimum velocity of $\gtrsim 4,500$
   \kms (i.e., for $\alpha_{\rm LOS} = 90^\circ$).

   We note here that such a ``high velocity'' for the scatterer does not
   introduce a significant time-delay for the reflected spectrum relative to
   the one that is directly observed.  For instance, a scatterer characterized
   by $v_{\rm rec} = 4,500$ \kms\ would be located on day 167 roughly $2.5$
   light-days from the center of the nebula.  For $\alpha_{\rm LOS} =
   90^\circ$, the time delay would thus be $\sim 2.5$ days; the maximum
   possible delay time of $\sim 5$ days would result for $\alpha_{\rm LOS} =
   180^\circ$.  During the nebular phase, SN spectra evolve very slowly, and so
   a few days of delay has negligible impact on the spectrum of the reflected
   light relative to that received directly.

 \item {\bf For a single external scatterer, the breadth of unblended line
   features in polarized flux is driven most strongly by the opening angle,
   with larger values serving to increase the breadth of the line.}  In
   Appendix~\ref{appendixb2}, we derived the absolute upper limit on the
   broadening of the H$\alpha$ line in the day 167 polarized flux spectrum of
   SN~2013ej to be 24\%. From our models, any opening angle $\gtrsim 40^\circ$
   for the scatterer yielded more than this amount of broadening in the
   polarized flux, thus placing an upper limit of $2\beta \lesssim 40^\circ$ on
   the external scatterer present in SN~2013ej.

 \item {\bf A bi-polar distribution of identical high-velocity scatterers is
   only consistent with the data for a narrow range of orientations.}  We also
   investigated the effect of multiple scatterers by performing simulations
   with a ``bi-polar'' distribution of identical clumps --- i.e., two clumps
   $180^\circ$ apart, a popular set-up for jet-like explosion
   mechanisms \citep{Utrobin17b,Izzo19}.  Similar to the results of
   \citet{Dessart21b}, we found that LOS angles of $\alpha_{\rm LOS} \lesssim
   60^\circ$ (for the approaching clump) produced a clear ``double peaked''
   profile in the polarized flux, decidedly inconsistent with observations of
   SN~2013ej.  Angles between $60^\circ \lesssim \alpha_{\rm LOS} \lesssim
   80^\circ$ produced line profiles which, while no longer double-peaked, were
   still unacceptably broadened relative to the constraints set on SN~2013ej.
   From this, we conclude that if a bi-polar distribution of identical
   scatterers is present in SN~2013ej, it is limited to inclinations within
   $\sim 10^\circ$ of the plane of the sky (i.e., $80^\circ \lesssim
   \alpha_{\rm LOS} \lesssim 100^\circ$).

\end{enumerate}


\end{document}